# Surface plasmons at single nanoholes in Au-films

L. Yin[1,3], V. K. Vlasko-Vlasov[1], A. Rydh[1], J. Pearson[1], U. Welp[1], S.-H. Chang[2,4], S. K. Gray[2], G. C. Schatz[4], D. E. Brown[3], C. W. Kimball[3]

[1]Material Sciences Division & [2]Chemistry Division, Argonne National Laboratory, Argonne, IL 60439

[3] Physics Department, Northern Illinois University, DeKalb, IL 60115

[4]Chemistry Department, Northwestern University, Evanston, IL 60208

The generation of surface plasmon polaritons (SPP's) at isolated nanoholes in 100 nm thick Au films is studied using near-field scanning optical microscopy (NSOM). Finite-difference time-domain calculations, some explicitly including a model of the NSOM tip, are used to interpret the results. We find the holes act as point-like sources of SPP's and demonstrate that the interference between the SPP and a directly transmitted wave allows for the determination of the wavelength, phase, and decay length of the SPP. The near-field intensity patterns can be manipulated by varying the angle and polarization of the incident beam.

Light manipulation on a sub-wavelength scale is desirable for nanophotonics applications. This could be achieved by developing nanostructures that transfer, turn, split, or tune light intensity via coupling to intrinsic electromagnetic (EM) modes in these structures. In this regard, surface plasmon polaritons (SPP's) are intrinsic EM modes of great interest [1,2]. SPP's localize and amplify light near metal surfaces and result, for example, in strong transmission enhancement through sub-wavelength holes in periodic structures [2,3].

We study SPP generation at *isolated nano-scale holes* in metal (Au) films, which could form building blocks of more complex photonic structures. We use near-field scanning optical microscopy (NSOM) to show that the holes act as point-like sources of SPP's. Interference with directly transmitted light through the film allows us to determine the wavelength, phase, and decay length of the SPP's. The SPP's propagate in directions determined by the polarization of the incident beam with an intensity decay length of $\delta \approx$ 1.0 $\mu$m. The resulting light patterns can be controlled by changing the polarization and angle of incidence of the illuminating beam. The experimental features are reproduced in finite-difference time-domain (FDTD) calculations [4]. Calculations with and without a probe model show that the NSOM images reflect the unperturbed Poynting vector normal to the surface.

The experiments are on 100 nm thick Au films sputtered onto 380-μm fused quartz substrates. Individual holes of 200 nm diameter are created by focused ion beam milling. A near-field scanning optical microscope (Aurora, TM) images the optical near field around the holes. The samples are illuminated from the substrate by a 532 nm laser beam (see Fig. 1e). The light intensity on the top surface is measured using Al-coated fiber probes with 50 to 80 nm apertures operating in collection mode at ~5 nm above the sample.

Fig. 1a shows the near-field optical pattern around a 200 nm hole in a normally illuminated Au film. The maximum intensity is over the hole. Lobes of enhanced intensity extend from either hole side over several μm along the direction of the incident polarization. The lobes have pronounced interference fringes with period ~475 nm, significantly shorter than the incident wavelength. The rotation of the polarization turns the lobes by the same angle (Fig.1b). Similar lobe patterns and polarization dependence, although without interference fringes, have been observed when SPP's are launched using local illumination through a NSOM tip [5].

The fringe period in Fig. 1 coincides with the SPP wavelength on top of a flat gold film,

$\lambda_{SPP} \approx \lambda_i \sqrt{\frac{Re(\varepsilon_m) + 1}{Re(\varepsilon_m)}}$ = 471 nm for incident light $\lambda_i$ = 532 nm and Au dielectric constant

$\varepsilon_m$ = -4.67 + i 2.42 [6]. Such a coincidence is at first not clear. For example, interference between SPP's launched from arrays of holes results in intensity modulations with period $\lambda_{SPP}/2$ [7]. The single pass attenuation through the thickness h = 100 nm of gold results in direct transmission of $\exp[-2\pi Im(\varepsilon_m^{1/2})h/\lambda_i]$ ~ 7% of the incident amplitude. Here, we show that the interference of this transmitted wave with SPP's on the surface accounts for the $\lambda_{SPP}$-fringing. Let the film be in the x-y plane between z = -h and 0, with the hole centered at x = y = 0. Assume x-polarized light propagating up from z = - ∞. Along x and near z = 0 we expect $E_x \approx A \exp(-|x|/2\delta) \exp(ik_{SPP}x) / |x|^{1/2}$ + B, $E_y \approx 0$, and $E_z \approx C$ $\exp(-|x|/2\delta) \exp(ik_{SPP}x) / |x|^{1/2}$, omitting the common $\exp(-i\omega_i t)$ terms. B in the 2nd part of $E_x$ is the directly transmitted amplitude. It is easy to see that $|E_x|^2$, the total intensity $|\mathbf{E}|^2$, and the normal component, $S_z$, of the time-averaged Poynting vector, $\mathbf{S} = (1/2)\,\text{Re}(\mathbf{E} \times \mathbf{H}^*)$, all contain an oscillatory term proportional to $\cos(k_{SPP}x)$, i.e. display interference fringes with period $\lambda_{SPP} = 2\pi/k_{SPP}$. While previously considered theoretically [8], this mechanism was not detected in earlier NSOM studies of individual nanoholes [9, 10]. If $p$ polarized light is incident at angle $\alpha$ to the film normal (Fig. 1c), the 2$^{nd}$ term in $E_x$ is $B\exp(ik_i x \sin\alpha)$ and the oscillatory term in the intensity becomes $\cos[(k_{SPP} - k_i \sin\alpha)x]$. In the +x direction the fringe period is $2\pi/(k_{SPP} - k_i\sin\alpha)$, which is larger than it is in the opposite (-x) side, $2\pi/(k_{SPP} + k_i\sin\alpha)$. We see such changes upon tilting the laser beam, as shown in Fig. 1c. The fringes have periods of 405 and 600 nm, respectively, in good agreement with the experimental tilt angle $\alpha \approx 13^0$. This asymmetry is even observed for $s$ polarization, Fig. 1d.

Fig. 2 shows experimental traces along the lines in Fig. 1a and 1c, plotted as $(I-I_0)|x|^{1/2}$, and fits according to the simple model above. $I_0$ is a constant off-set measured at large distance from the hole. Except in the hole vicinity, where additional light intensity due to waves passing directly though the hole arises, the data are described well. In addition to the determination of the fringe periods for normal (Fig. 2a) and inclined (Fig. 2b) incidence, an intensity decay length of $\delta \approx 1.0$ μm is found. The nature of the NSOM signal is less obvious because both $|E_x|^2$ and $|\mathbf{E}|^2$, within the simple model above, have the same parametric form and cannot be distinguished with our fits. The FDTD simulations will address this issue.

2D-light patterns on the film surface are obtained from FDTD simulations [4,11] with a 100 nm Au layer and total system sizes up to 7.5 x 7.5 x 1 μm$^3$, with 4 nm grid resolution. Owing to the large system size, a parallel FDTD implementation is used.

Fields near grid boundaries are absorbed with a uniaxial perfect matching layer (UPML) [4]. A Drude model fit to the refractive index data of Au in the 530-nm region is employed for the film and implemented as in Ref. [12]. Fourier transforms of the results yield the electric and magnetic fields. Fig. 1f displays an image map of $S_z$, evaluated at ~ 6 nm above the surface. It is in qualitative agreement with the data in Fig. 1a, including fringes separated by $\lambda_{SPP}$. We also find (not shown here) that $|E_x|^2$ is qualitatively similar to $S_z$, whereas $|\mathbf{E}|^2$ shows some structures in addition to the $\lambda_{SPP}$-fringes. Also, calculations for a film with no hole allow us to infer directly transmitted fields and, upon subtraction from our full film/hole fields we find a pattern without fringes, further verifying the role of interference.

Fig. 3 shows (thin line) a horizontal cut (y = 0) of the FDTD result of Fig. 1f. While it qualitatively agrees with the corresponding experimental trace (thick curve), note negative energy flow near the nanohole edge. This is consistent with energy losses due to SPP creation at hole edges, which is, however, not seen experimentally. To understand the NSOM probe effect [13], we introduce a cone-shaped fiber tip with Al coating above the film (inset, Fig. 3). The tip aperture diameter is 80 nm, and opens up to a cylinder of diameter 240 nm. The total height of the tip structure is 0.6 μm and it is truncated by the UPML. A large model tip is necessary in order for it to support propagating modes. Fig. 3 shows the surface integral of the Poynting vector over the cylinder top (symbols), which is a measure of the NSOM signal. Note that negative energy flow near the hole edges has now disappeared. Fig. 3 shows that the tip calculations preserve the structure of the calculations without tip and agree reasonably well with experiment. The experimental curve is shifted by 0.1 μm to the left in order to line up the fringes. The reason for this slight discrepancy between experiment and theory could be related to irregularities in the experimental tip such as protrusions or uneven Al-cladding.

To conclude, sub-wavelength holes in thin metal films act as a point-like SPP sources at normal or near normal incidence.  This source can be manipulated by the conditions of illumination.  The results are accounted for in FDTD calculations, which also show that while the presence of the NSOM tip can perturb the EM fields, the overall unperturbed patterns are preserved in the estimated power going up the tip.  We are currently studying different geometrical configurations of such holes that can generate qualitatively new light patterns with potential for future nanophotonics applications.

The work at ANL was supported by the US Department of Energy, Basic Energy Sciences, under contract W-31-109-ENG-38, and at NIU by the State of Illinois under HECA and the US Dept. of Education.  Work with the FIB was carried out in the Center for Microanalysis of Materials, University of Illinois, which is partly supported by the US Department of Energy under grant No. DEFG02-91-ER45439.  GCS was supported by the US Department of Energy under grant No. DEFG02-03-ER15487.  The calculations were performed at the High Performance Computing Research Facility, Mathematics and Computer Science Division, Argonne National Laboratory.

**Figure Captions**

Fig. 1  Near-field optical images with normal illumination and horizontal (a) and vertical (b) polarization. The period of fringes is about 475 nm. (c) and (d): Near-field optical images with inclined illumination. The polarization directions are shown as arrows. The lines in (a) and (c) indicate the location of the experimental traces shown in Fig. 2. (e): Experimental set-up. (f) FDTD calculation of the normal component of the Poynting vector. All frames are 7.5 μm x 7.5 μm.

Fig. 2  Experimental traces (solid) along the lines indicated in Figs. 1a and 1c, and fits (dashed) according to the SPP point-source model.

Fig. 3  The (y = 0) x-axis trace through Fig. 1f (thin line), FDTD calculations of the tip effect (open squares), and experimental trace (thicker solid line). Each square is from a separate FDTD calculation with the tip located at different points on the x-axis. The experimental trace was shifted by 0.1 μm to the left in order to align the fringes (see text).

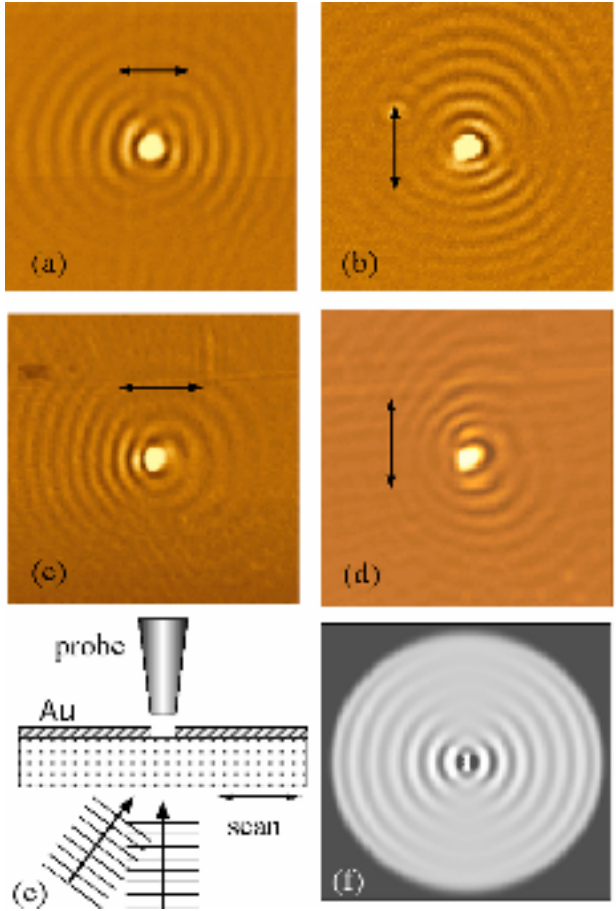

Fig. 1        L. Yin et al

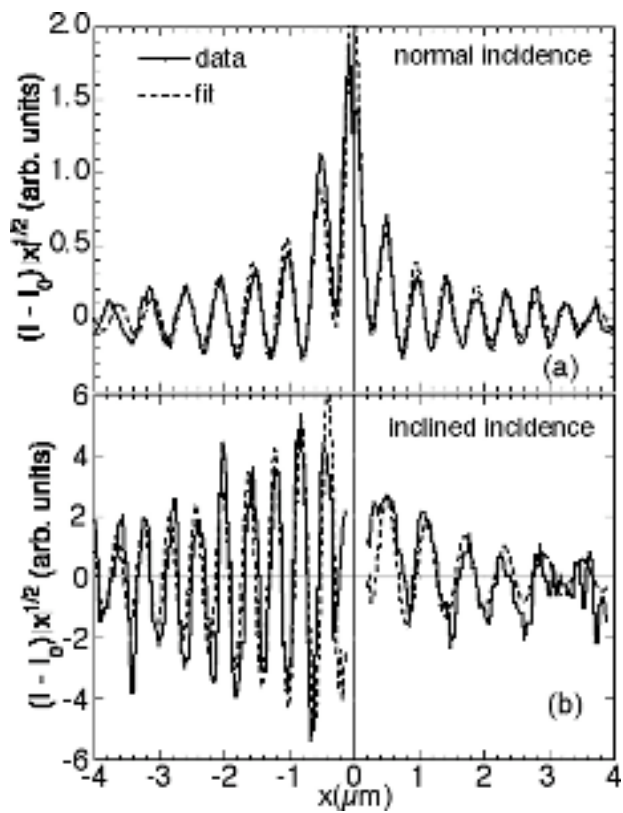

Fig. 2 L. Yin et al.

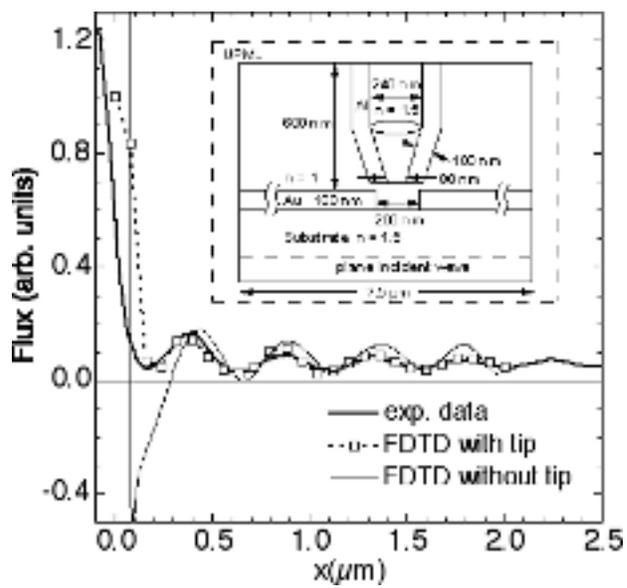

Fig. 3 L. Yin et al.